\documentclass[preprint,12pt,authoryear]{elsarticle}

\usepackage{amssymb}
\usepackage{amsmath}
\journal{Nuclear Instruments and Methods in Physics Research Section A: Accelerators, Spectrometers, Detectors and Associated Equipment}

\begin{document}

\begin{frontmatter}



\title{ANISSA: Advanced Neutron Imaging for Solid-State batteries in Action} %% Article title


 %% Author name
\author[label1]{Oriol Sans-Planell}
\author[label1]{Nikolay Kardjilov}
\author[label1]{Ingo Manke}
\author[label1]{Gitanjali Gitanjali}
\author[label2]{Martin Lange}
\author[label3]{Eva Schlautmann}
\author[label4,label5]{Alessandro Tengattini}
\author[label6]{Stephen Hall}
\author[label6]{Philip Vestin}
\author[label7]{Qaphelani Ngulube}
\author[label8]{Robin Woracek}
\author[label2,label3]{Wolfgang G. Zeier}
\author[label7]{Kristina Edström}

\affiliation[label1]{organization={Helmholtz-Zentrum Berlin},%Department and Organization
            addressline={Hahn-Meitner Platz 1}, 
            city={Berlin},
            postcode={14109}, 
            country={Germany}}
\affiliation[label2]{organization={Institute of Energy Materials and Devices (IMD) IMD-4: Helmholtz-               Institut Münster, Forschungszentrum Jülich},%Department and Organization
            addressline={Corrensstrasse 48}, 
            city={Münster},
            postcode={48149},
            country={Germany}}
\affiliation[label3]{organization={Institute of Inorganic and Analytical Chemistry, University of Münster},%Department and Organization
            addressline={Corrensstrasse 28/30}, 
            city={Münster},
            postcode={48149},
            country={Germany}}
\affiliation[label4]{organization={Institut Laue-Langevin},%Department and Organization
            addressline={Av. des Martyrs 71}, 
            city={Grenoble},
            postcode={38000},
            country={France}}
\affiliation[label5]{organization={Université Grenoble-Alpes},%Department and Organization
            addressline={Av. Centrale}, 
            city={Saint-Martin-d'Hères},
            postcode={38400},
            country={France}}
\affiliation[label6]{organization={University of Lund},%Department and Organization
            addressline={Professorsgatan 1}, 
            city={Lund},
            postcode={223 64},
            country={Sweden}}
\affiliation[label7]{organization={University of Uppsala},%Department and Organization
            addressline={Angstrom Laboratory, Box 538}, 
            city={Uppsala},
            postcode={751 21},
            country={Sweden}}
\affiliation[label8]{organization={European Spallation Source},%Department and Organization
            addressline={Partikelgatan 2}, 
            city={Lund},
            postcode={224 84},
            country={Sweden}}

%% Abstract
\begin{abstract}
The development of high-energy density solid-state batteries is critical for the achievement of carbon neutrality goals and the advancement of clean energy. Still, the fundamental understanding of lithium transport mechanisms and degradation processes remains limited. Current characterisation methods face significant challenges in studying these complex systems, particularly due to the difficulty of detecting lithium dynamics in three-dimensional battery architectures in \textit{operando} conditions. Here we present the ANISSA (Advanced Neutron Imaging for Solid-State batteries in Action) project, an integrated experimental framework combining high-resolution neutron and X-ray imaging techniques to research coupled electro-chemo-mechanical processes in lithium-based energy storage systems.
\end{abstract}

\begin{graphicalabstract}
\includegraphics[width=\textwidth]{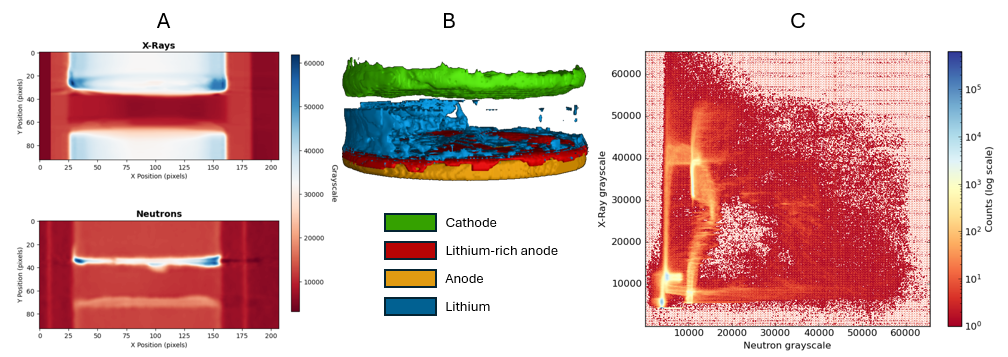}%
\end{graphicalabstract}

\begin{highlights}
\item The new project ANISSA will bring new insights into electro-chemo-mechanical processes within solid-state batteries.
\item Multimodal high-resolution 5D imaging is the best non-destructive frontier method to study solid-state batteries.
\end{highlights}

%% Keywords

\end{frontmatter}


\section{Introduction}
The "Advanced Neutron Imaging for Solid-State batteries in Action" (ANISSA) project has been funded by the German BMBF's \textit{Batterieforschung} funding scheme. \\
The project will exploit recent advances in in-situ and operando experimental characterisation of battery operation with neutrons \citep{siegel2011neutron,bradbury2023visualizing,bradbury2023visualizing2} and X-rays \citep{sun2017complementary}, together with advances in modelling and simulation \citep{scharf2022bridging}, to enhance understanding of coupled electro-chemo-mechanical processes in lithium-based energy storage systems.\\
The partners involved in the ANISSA project are the following: The University of Münster and University of Uppsala, who will be leading the development of solid-state batteries, the University of Lund and the Helmholtz-Zentrum Berlin who will provide the experimental and analytical expertise, and the European Spallation Source (ESS) in Sweden and the Institute Laue-Langevin (ILL) in France, who will participate as leading large-scale facilities providing the means for the experimental characterization of the batteries.\\
The ANISSA project will accelerate the development of instrumentation for early science at ESS concerning a major challenge of exceptional societal impact: the realisation of high-energy density solid-state battery technologies.

\section{Project description}
\label{sec1}
%% Labels are used to cross-reference an item using \ref command.
Lithium-based batteries constitute a key technology in the European green deal and vision to be the first climate-neutral continent, before 2050 \citep{arbabzadeh2019role,zeng2022battery}. Such batteries are expected to be vital for electromobility \citep{von2024future}, load-levelling of intermittent energy sources \citep{jiang2025battery} and to power the growing portable electronics and internet-of-things sectors \citep{krishna2024advanced}. Batteries are essential for achieving the carbon-neutral strategies of many countries, contributing to reduced air pollution and mitigating climate change. In this quest, there is an urgent need to accelerate the identification and realisation of new ultra-high-performance battery materials and battery concepts that can surpass the performance of the current best rechargeable Li-ion batteries, which suffer from limited energy storage and safety that limit their versatility. Solid-state batteries utilising metallic lithium as the negative electrode and high-capacity, high-voltage materials as the positive electrode are among the most challenging battery chemistries, but have the potential to double storage capacity with improved safety \citep{ye2024fast,wang2024lithium,lee2020high}. For such batteries to be realised, many basic research questions need to be solved to avoid drastic capacity reduction and to mitigate potentially dangerous short-circuiting. Answering these questions requires improved understanding of, e.g.: three-dimensional diffusion processes in the complex materials; formation of passive layers between battery components; and dendrite formation from the lithium metal. \\
\begin{figure*}
\includegraphics[width=\textwidth]{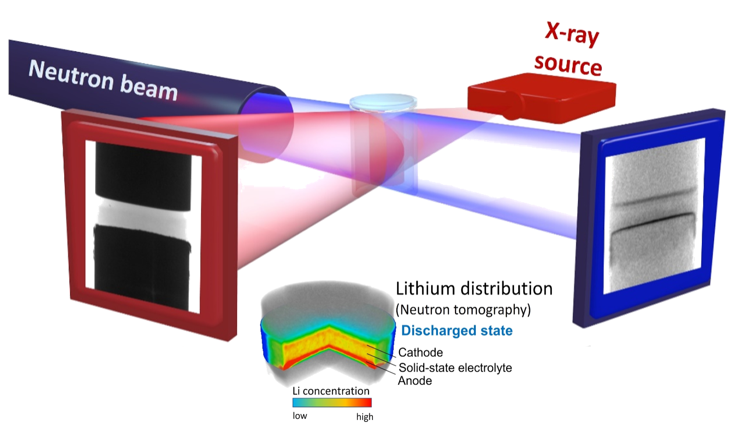}%
\caption{\label{fig:NeXT}Model of the geometry present at NeXT and ODIN for the dual-mode tomography. The two radiation beams are installed at a 90-degree angle, and can take the tomography simultaneously, allowing for \textit{operando} measurements of the samples. On the bottom, a highlight of the segmentation possible of a neutron tomography performed on a solid-state battery, showing the distribution of lithium in the electrolyte in a discharged state.} 
\end{figure*}
\begin{figure*}
\includegraphics[width=\textwidth]{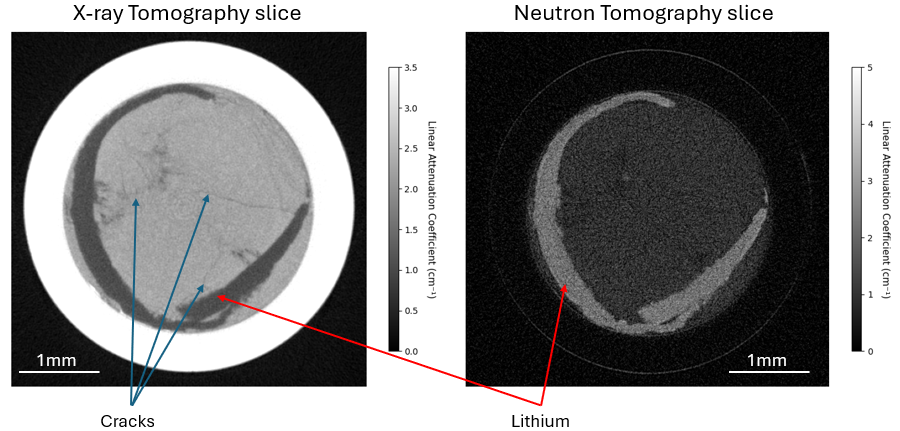}%
\caption{\label{fig:Lithium}Comparison between two slices of a tomography taken of the same solid-state battery at NeXT, ILL, with X-rays (left) and cold neutrons (right). The X-rays show very defined cracks in the electrolyte, while the neutrons offer a much better contrast on the Lithium present in the volume.} 
\end{figure*}
To unravel the complex phenomena described above for conventional Li-ion batteries, two- or three-dimensional (2D/3D) imaging techniques can be exploited to provide locally resolved information \citep{ziesche20204d}. X-ray imaging has become a standard method for 2D and 3D imaging in many disciplines, but has some major limitations for battery studies, primarily due to its low sensitivity to light elements like lithium \citep{lu20203d}. Meanwhile, neutron imaging has rapidly grown in importance as a non-destructive analytical tool in many disciplines due to enhanced neutron sources and detectors enabling higher spatial and temporal resolutions \citep{magnier2021tomography}. Furthermore, the development of alternative neutron imaging modalities, exploiting scattering and wavelength dependence, makes neutron imaging increasingly attractive. \\
ANISSA focuses on the application of dual-modality, multi-spectral, high-resolution and high-speed imaging techniques to reveal the distribution and movement of the key component, lithium, in solid-state battery material components and in complete devices. Advanced facilities, such as NeXT at ILL, offer these multimodal tomography applications as a standard technique, represented in Figure \ref{fig:NeXT}. \\
For battery studies, neutron imaging can overcome a number of the limitations of X-ray imaging, primarily due to the high sensitivity to lithium, which enables direct detection of lithium diffusion. This can be visualised in Figure \ref{fig:Lithium}, where the attenuation produced my Lithium generates a very strong contrast with the surrounding material, especially when compared to the X-rays, in which the lithium can barely be distinguished from the air. This allows the visualisation of Li-ion migration and subsequent identification of areas of reduced activity responsible for battery capacity fading. During dis-/charge, dynamic 3D measurement of the change in Li concentration can be obtained by measuring spatio-temporal changes in neutron attenuation coefficients over a battery volume. Furthermore, neutrons can also be used to distinguish between isotopes, such that Li exchange processes inside a battery can be followed by doping of different battery components \citep{bradbury2023visualizing}. \\ 
The advanced techniques described are versatile enough to contribute to analysing a large variety of battery chemistries, including the dominant Li-ion types, thereby creating a long-lasting benefit and impact for future research and developments. Three major imaging developments are targeted:
\begin{enumerate}
    \item A combined high-resolution X-ray and neutron imaging methodology with much improved spatial resolutions to link Li mobility and distributions to morphology changes in solid-state batteries, and correlate to its electrochemical performance during operation. This will be the first dual-mode installation with high spatial resolutions of a few µm, a necessity to study individual battery particles.
    \item A novel isotope mapping technique in combination with 5D multispectral imaging (3D space plus time plus energy selective) capabilities for exploiting the contrast variations between $^6$Li, $^7$Li, $^{nat}$Li to provide important information about the dynamic properties of the lithium diffusion during battery operation.
    \item An optimised cell design for multi-modal and multi-spectral correlative imaging in a combination with a dedicated portable in-situ sample environment for solid-state battery operation and conditioning for use at ESS as soon as it comes into operation, and with first demonstrations at ILL.
\end{enumerate}
Each of these developments will present an unrivalled progression from the current state-of-the-art, while the combined and dedicated efforts through the ANISSA project will provide an unprecedented characterisation tool for investigation of the Li exchange processes inside solid-state batteries with high spatial and temporal resolution.

\section{Facilities}
\subsection{ODIN at ESS}
The Optical and Diffraction Imaging with Neutrons (ODIN), from the European Spallation Source (ESS), will be the focus for the technique developments of this project and will also be on the initial batch of beamlines to come into operation at ESS. It is expected that ANISSA will represent some of the first results yielded by the facility. \\
ODIN is a multi-purpose imaging beamline engineered to provide single-digit $\mu m$-range image resolution with very high neutron flux, coupled with state-of-the-art event-mode detector technology. The instrument is being built with a complex system of choppers which allows for constant wavelength resolution over the wide available spectrum. The time-of-flight capabilities enable advanced neutron imaging techniques, such as polarised neutrons, grating interferometry and Bragg-edge imaging. This environment is prime for the study of solid-state batteries, in particular with the ANISSA approach of lithium enrichment for the \textit{operando} multimodal studies. \\
At the time of writing, ESS is expected to begin operation in 2027. Until then, the development of the project is expected to take place at the Institute Laue-Langevin, in France.
\subsection{NeXT at ILL}
The Neutrons and X-Rays Tomography (NeXT) beamline, at the Institute Laue-Langevin, in Grenoble, France, is a well-established, state-of-the-art neutron facility which offers the most powerful cold neutron beam in the world \citep{tengattini2020next}. The high flux offered by NeXT allows for very high-resolution measurements, down to 3$\mu$m. Due to the high flux, NeXT also provides the possibility for multimodal \textit{operando} measurements of small samples, such as solid-state batteries. These samples in particular can greatly benefit of the correlative information of this technique, as shown in Figure \ref{fig:bivariate_histogram}.\\
In addition to the 4D imaging, NeXT will also allow for \textit{operando} measurements, both in radiography and tomography, using advanced imaging techniques, such as polarised neutron imaging, grating interferometry or monochromatic imaging by means of the double crystal monochromator setup.

\section{Scientific advancement and results}

Research on solid state batteries has been revived during recent years towards doubling practical capacity compared to the successful Li-ion batteries, which today dominate the market for both portable electronics and the transport sector. We have a strong driving force to obtain new scientific insights, which in the best of worlds could be transferred to other more applied research projects, as new innovative battery solutions. The significance of the project will be unprecedented in terms of new knowledge about subtle Li-ion transport reactions in solids, especially at interfaces and interphase development in solid-state batteries. The development of advanced imaging of battery operation, combining attenuation of X-rays and neutrons, will provide a key, currently missing, component in the toolbox for characterising battery materials and battery concepts.
\begin{figure*}
\includegraphics[width=\textwidth]{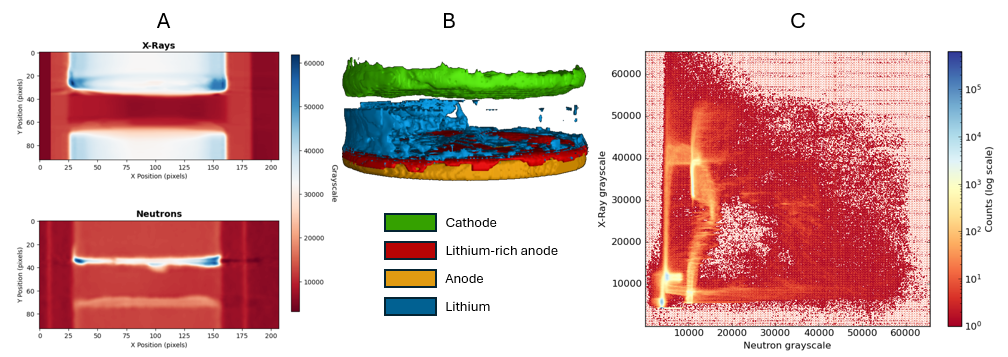}%
\caption{\label{fig:bivariate_histogram}A: visualisation of two slices of a tomogram taken by X-rays (top) and neutrons (bottom) at the ILL beamline using a high-resolution set-up. B: Segmentation of an NCM-811 solid-state battery by combining the information from two registered neutron and x-ray tomograms. C: Representation of the multimodal histogram showing the combined information from the two registered volumes.}%
\end{figure*}
As shown in Figure \ref{fig:bivariate_histogram}, the combination of information provided by both high-resolution neutrons and x-rays will allow for a detailed study of the phenomena occurring in \textit{operando} conditions inside the solid-state battery. \\
The experimental arrangement for the multimodal measurements is determined by the different beam geometries. The neutron beam is considered parallel, contrary to the X-rays, which have a conical shape. The resolution in case of neutron setup $R_n$ is determined by the collimation ratio $L/D$ ($L$ - collimator length, $D$ - collimator aperture) and the distance between the sample and detector $l$ as $R_n=l/(\frac{L}{D})$. The X-ray resolution $R_x$ depends on the magnification ratio $M=SD/SO$ ($SD$ - source-detector and $SO$ - source-object distances) as $R_x=P_x/M$, where $P_x$ is the pixel size of the X-ray detector. In this way, we have two border conditions for simultaneous high-resolution imaging with X-rays and neutrons:
\begin{enumerate}
    \item The sample should be close to the neutron detector, keeping the distance $l$ small.
    \item The sample should be close to the X-ray source, resulting in short SO.
\end{enumerate}  
\begin{figure*} \centering
\includegraphics[width=10cm]{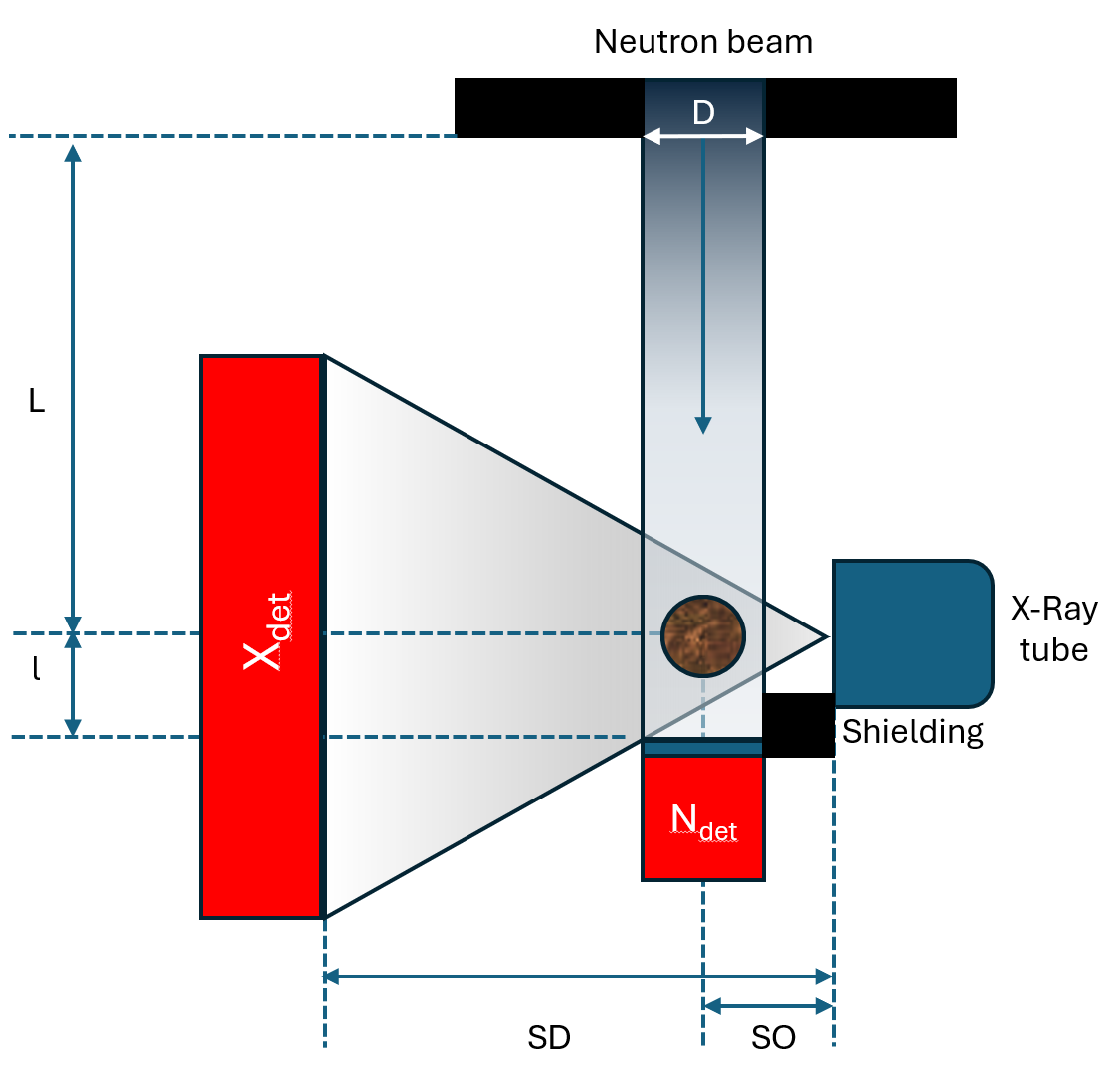}%
\caption{\label{fig:bivariate_geometry}Scheme showing the geometrical elements that compose the multimodal imaging set-up, not in scale. On the vertical axis, the neutron beam is constrained by a pin-hole with an aperture $D$, which will define the collimation ratio $L/D$, where $L$ is the distance between the pin-hole and the sample. The sample is placed at a distance $l$ from the detector, represented by $N_{det}$. Perpendicular to the neutron beam there is the X-ray set-up, with the cone beam geometry produced by the X-ray tube. The sample in this case is at a distance $SO$ from the source, and its projection is magnified to the detector $X_{det}$ placed at a distance $SD$ from the source. There must be a shielding element protecting the neutron scintillator from the influence of the X-ray beam.}%
\end{figure*}
On the other hand, the X-ray beam should not illuminate the neutron detector because of its sensitivity to X-rays and the neutron beam should not irradiate the X-ray source to prevent radiation damage to the electronics. Combining all these constraints, a complex arrangement arises, which provides some limitations for the aimed spatial resolution. For example, the SO distance is determined by the width of the neutron detector if the sample is placed very close to it. In practical cases, the SO distance is in the range of 10 cm, which results in a magnification range of 5 for a typical SD distance of 50 cm. For achieving spatial resolution below 5 µm, one of the goals of the ANISSA project, the pixel size of the X-ray detector should be smaller than 25 µm. These criteria determined the selection of the optimal X-ray detector based on a sCMOS X-ray camera, allowing for high-resolution direct phosphor imaging.\\
The new equipment and approaches installed at ESS, plus an optimised electrochemical operando cell, will lead to solutions to many fundamental unanswered questions currently hindering the success of solid-state batteries. The scientific novelties concerning Li batteries in the project will be: 
\begin{enumerate}
    \item Imaging to provide new understanding of the Li-ion transport properties through the electrolyte (ceramics, polymer electrolyte and composite-electrolytes).
    \item New insights into ion transport through (evolving) interfaces, which is a challenging issue for many solid electrolytes (instability vs. metallic Li or a high voltage cathode).
    \item New methods to create battery materials and for interface engineering to mitigate Li dendrite formation.
\end{enumerate}
There are multiple innovative aspects concerning instrumentation and experimental methodology, which will significantly push the conventional boundaries. The X-ray setup will be the first one that will offer high resolution (below 5 µm) in combination with neutron imaging and will be the first such installation at a pulsed neutron source. Being on a pulsed source (and by using the new monochromatic option at the ILL during the first part of the project) will enable much improved quantification, by correcting for systematic biases introduced by scattered neutrons \citep{boillat2018chasing}, and enable spectral neutron imaging \citep{tran2021spectral}, yielding information that is otherwise only attainable from neutron diffraction (ND) \citep{woracek2018diffraction}, quasi-elastic neutron scattering (QENS) and inelastic neutron scattering (INS) \citep{siegwart2019distinction}. Imaging with the ND, QENS and INS as contrast modalities is very recent, and ANISSA will present ideal conditions to elevate the methodologies at a new level. 

\section{Conclusions}
The ANISSA project will offer unprecedented insights into the electro-chemo-mechanical processes within solid-state batteries. The technological innovation will span over multiple interdisciplinary applications, with developments in multimodal high-resolution \textit{operando} 5D imaging, battery characterisation and data analysis, bridging current gaps in scientific knowledge. \\
Through the advanced materials design, the project will advance the current frontier of knowledge regarding the transport limitations on the electrolyte, the interface stability between the electrodes and electrolytes and dendrite formation. This research will establish a new paradigm for battery characterisation that can be transferred to other systems beyond the solid-state. The results from this project will impact the development of next-generation battery concepts, becoming a cornerstone for energy storage developments.\\

\section*{Acknowledgment}
We would like to thank the German Federal Ministry of Research,
Technology and Space (BMFTR, BMBF) for funding the project ANISSA (proj no. 05K2022 with sub-project 05K22CBA). Parts of this work were supported by the BMFTR/BMBF project LISZUBA (project no. 03XP0115C). Furthermore, we would like to thank the Helmholtz Association for supporting the project.
%\bibliographystyle{elsarticle-harv} 
%\bibliography{ANISSA_Nonames}



\end{document}